\documentclass[a4paper,11pt]{article}
\usepackage{pos}

\title{Results on spin sum rules and polarizabilities at low $Q^2$}

\author*[a]{A. Deur}

\affiliation[a]{Jefferson Lab,\\
12000 Jefferson Avenue, Newport News, VA 23606, USA}

\emailAdd{deurpam@jlab.org}

\abstract{We report on recently published experimental results on spin sum
rules, and particularly on the generalized spin polarizabilities
$\gamma_0(Q^2)$ (for both the proton and neutron) and $\delta_\mathrm{LT}(Q^2)$
(for the neutron).
The data were taken at Jefferson Lab in Hall A by experiment E97110 (neutron) and in Hall B by experiments 
E03006 and E05111 (proton and deuteron, respectively). The experiments covered
the very low $Q^2$ domain, down to $Q^2 \simeq 0.02$ GeV$^2$. This is well into the domain where Chiral
Effective Field Theory ($\chi$EFT) predictions should be valid. 
Some measured observables agree with the state-of-the-art $\chi$EFT 
predictions but others are in tension, including $\delta_\mathrm{LT}^n(Q^2)$ 
which $\chi$EFT prediction was expected to be
robust. This suggests that $\chi$EFT does not yet consistently describe
nucleon spin observables, even at the very low $Q^2$ covered by
the experiments.
}

\FullConference{%
  The 10th International Workshop on Chiral Dynamics - CD2021\\
  15-19 November 2021\\
  Online
  }

\begin{document}
\maketitle

\section{Introduction and background}

Chiral effective field theory ($\chi$EFT) is the leading effective theory describing the first level of complexity emerging 
from the Standard Model, {\it viz} how the fundamental quarks and gluons produce hadronic and nuclear phenomena.
As such, $\chi$EFT is a crucial component of our global understanding of Nature. It has been very successful in 
explaining hadronic and nuclear physics~\cite{Bernard:1995dp}. Yet, its description of 
nucleon spin structure remains imperfect. Table~\ref{tab:comp_old} (adapted from~\cite{Deur:2018roz}) lists nucleon spin observables measured in the late 1990s and early 2000s at Jefferson Lab (JLab) 
and shows how well the $\chi$EFT predictions available at the time described them.
(What the observables mean is unimportant here. Their definitions will be given in latter sections.) 
\begin{table*}[!h]
\resizebox{0.999\textwidth}{!}{%
\begin{tabular}{|c|c|c|c|c|c|c|c|c|c|} \hline
~ & $\Gamma_1^p$~[3, 4]    &  $\Gamma_1^n$~[4, 5] & $\Gamma_1^{p-n}~[6, 7]$  & $\Gamma_1^{p+n}~[4, 7]$ & $\gamma_0^{p}~[4]$ &$\gamma_0^{n}~[8]$ &$\gamma_0^{p-n}~[7]$ &$\gamma_0^{p+n}~[4, 7]$ &  $\delta_\mathrm{LT}^n~[8]$ \\ \hline 
Ji {\it et al.}~\cite{Ji:1999pd} & {\bf \color{red}{X}} & {\bf \color{red}{X}} & {\bf \color{blue}{A}} & {\bf \color{red}{X}} & {\bf -} & {\bf -}  & {\bf -}  & {\bf -}  & {\bf -}\\ \hline
Bernard {\it et al.}~\cite{Bernard:2002bs} & {\bf \color{red}{X}} & {\bf \color{red}{X}} & {\bf \color{blue}{A}} & {\bf \color{red}{X}} & {\bf \color{red}{X}} & {\bf \color{blue}{A}}& {\bf \color{red}{X}}  & {\bf \color{red}{X}}  &{\bf \color{red}{X}}\\ \hline 
Kao {\it et al.}~\cite{Kao:2002cp} & {\bf -}  &{\bf -}    &{\bf -}   & {\bf \color{red}{X}} & {\bf \color{blue}{A}}& {\bf \color{red}{X}}  & {\bf \color{red}{X}} & {\bf -}  &{\bf \color{red}{X}}\\ \hline 
\end{tabular}}
\vspace{-3mm}
\caption{ Nucleon spin observables measured at JLab by experiments E94010 and EG1~\cite{Fatemi:2003yh, Prok:2008ev, Amarian:2002ar, Deur:2004ti, Deur:2008ej, Amarian:2004yf}
compared to early predictions from $\chi$EFT~\cite{Ji:1999pd, Bernard:2002bs, Kao:2002cp}.
The  {\bf \color{blue}{A}} ({\bf \color{red}{X}}) letter indicates that data and prediction agree (disagree) over 
the range $0 \leq Q^2  \lesssim 0.1$ GeV$^2$. The {\bf -}  indicates that no prediction was available at the time.
}
\label{tab:comp_old} 
\end{table*}
The table shows that the early $\chi$EFT predictions were in tension with the data more often than not. 
Particularly puzzling was the discrepancy for the spin polarizability $\delta_\mathrm{LT}^n$ because its $\chi$EFT 
prediction was expected 
to be robust owing to the suppression of the $\Delta_{1232}$ resonance contribution to $\delta_\mathrm{LT}$. 
This contribution is difficult to account for and was either not included in the early 
predictions~\cite{Ji:1999pd, Kao:2002cp}, or approximately included phenomenologically~\cite{Bernard:2002bs}. 
Was the origin of the discrepancy a $\chi$EFT calculation problem, maybe with the $\Delta_{1232}$? 
Or was it because the data were not at low enough $Q^2$ to reach the $\chi$EFT applicability domain? 
To answer these questions, refined $\chi$EFT calculations with improved expansion schemes and including the $\Delta_{1232}$ contribution 
were undertaken~\cite{Bernard:2012hb, Lensky:2014dda} and new experiments reaching well into the $\chi$EFT applicability 
domain were performed.
Here, we present results from that experimental program~\cite{Adhikari:2017wox, Zheng:2021yrn, Sulkosky:2019zmn, Sulkosky:2021qmh, Deur:2021klh}.

\section{Experimental method}
The observables listed in Table~\ref{tab:comp_old} are measured with inclusive inelastic lepton scattering in which
a lepton (for JLab, an electron) of momentum $p$ scatters off a nucleon or nucleus at rest in the laboratory frame. 
The lepton transfers a momentum $q=(\nu, \vec q)$ to the nucleon/nucleus whose fragments are not detected. 
Here, we will work within the one-photon exchange approximation, where $Q^2 \equiv -q^2 >0$ quantifies 
how virtual the exchanged photon is. 
The experiments discussed here, 
E97110~\cite{Sulkosky:2019zmn}, E03006~\cite{Zheng:2021yrn} and E05111~\cite{Adhikari:2017wox}
 were performed at JLab, a facility located in Newport~News, Virginia USA,
that accelerates a continuous electron beam to energies up to 12 GeV. The beam polarization for an experiment is typically $\sim$$ 85\%$.
Up to 200 $\mu A$ of beam can be circulated. It supplies four experimental halls, A, B C and D equipped either with high resolution
(A and C) or large acceptance (B and D) spectrometers. E97110 occured in Hall A and  E03006/E05111 (commonly referred to as Experimental Group EG4) in Hall B during the 6 GeV
era of JLab, before its upgrade to 12 GeV. 

The observables in Table~\ref{tab:comp_old} are obtained by integrating  over $\nu$ the nucleon polarized structure functions 
$g_1$ and $g_2$~\cite{Deur:2018roz}. To reach low $Q^2$ while keeping the wide  $\nu$ range 
necessary for the integration, a high-energy  beam (here up to 4.4 GeV) is needed  and the scattered electrons 
must be detected at small angles
(here, down to about $6^\circ$). These angles were reached in Hall A thanks to a new horizontally-bending dipole magnet 
placed in front of the spectrometer~\cite{septum}. In Hall B, a dedicated Cherenkov Counter optimized for high efficiency
at small angle was added to one of the six sectors (otherwise identical)  of the spectrometer~\cite{Adhikari:2017wox}. 
In addition, the Hall B target was moved 1~m upstream of its usual position and the spectrometer magnetic field was set 
to bent the electrons outward. 
E97110 studied the spin structures of the neutron and $^3$He  thanks to the Hall A  polarized $^3$He target, using
both its longitudinal and  transverse polarization capabilities. In particular, the latter allowed to measure $g_2$, which is crucial to form $\delta_\mathrm{LT}$. 
E03006/E05111  studied the proton, deuteron and neutron spin structures with the Hall B longitudinally 
polarized ammonia  (NH$_3$ or ND$_3$).

\section{Generalized spin polarizabilities}
Polarizabilities encode the second order reaction of a body subjected to an electromagnetic field, 
e.g. the reaction of a nucleon probed
by a low energy photon~\cite{Hagelstein:2015egb}. The complete  reaction is described by two Compton scattering amplitudes, $f_1$ (spin-independent) 
and $f_2$ (spin-dependent). Considering for now real photons ($Q^2=0$), one can expand $f_1$ 
and $f_2$ in term of $\nu$:
\begin{eqnarray}
f_1(\nu)=-\frac{\alpha}{M}+\big(\alpha_E+\beta_M \big)\nu^2+\mathcal{O}(\nu^4), \nonumber  \\
f_2(\nu)=-\frac{\alpha\kappa^2}{2M^2}\nu+\gamma_0\nu^3+\mathcal{O}(\nu^5), \nonumber
\end{eqnarray}
where $\alpha$ is the electromagnetic coupling, $M$ is the nucleon mass, $\kappa$ its anomalous magnetic moment, 
  $\alpha_E$ and $\beta_M$ are respectively the electric and magnetic polarizabilities, 
and $\gamma_0$ is the forward spin polarizability. The first term  in the equations ($\propto \alpha$) represents 
the purely elastic reaction expected from a perfectly rigid (or pointlike) object. 
The second term defines the polarizabilities and reflects the deformation of the object, i.e. its internal rearrangement. 
For virtual photons ($Q^2 \neq 0$), the polarizabilities acquire a $Q^2$-dependence --~they are then named {\it generalized polarizabilities}~-- and because virtual photons have a longitudinal spin component, 
the Longitudinal-Transverse polarizability $\delta_\mathrm{LT}$ appears. 

It is not known how to measure directly generalized spin polarizabilities. Instead, they are measured indirectly 
using the sum rules~\cite{GellMann:1954db}:
\begin{eqnarray}
\gamma_0(Q^2) =   \frac{16\alpha M^2}{Q^{6}}\int_0^{x_0}x^2\Bigl[g_1(x,Q^2)-\frac{4M^2}{Q^2}x^2g_2(x,Q^2)\Bigr]dx,
\label{eq:gamma_0}
\end{eqnarray}
\begin{eqnarray}
\delta_\mathrm{LT}(Q^2)  =   \frac{16\alpha M^2}{Q^{6}}\int_0^{x_0}x^2\Bigl[g_1(x,Q^2)+g_2(x,Q^2)\Bigr]dx,
\label{eq:delta_LT SR}
\end{eqnarray}
where $x=Q^2/2pq$ and $x_0$ is the inelastic threshold.

The function $g_1$ (and $g_2$ for E97-110) is measured from $x_0$ to a minimum non-zero $x$, since reaching 
$x=0$ with non-zero $Q^2$ requires infinite beam energy. $\gamma_0(Q^2)$ and $\delta_\mathrm{LT}(Q^2)$
are then obtained by integrating these measurements according to Eqs.~(\ref{eq:gamma_0}) and (\ref{eq:delta_LT SR})
 and using a parameterization~\cite{Adhikari:2017wox} to estimate the missing low-$x$ contribution.

\section{Experimental results on the generalized spin polarizabilities $\gamma_0(Q^2)$ and $\delta_\mathrm{LT}^n(Q^2)$}
\begin{figure}[!h]
  \centering
    \includegraphics[width=0.415\textwidth]{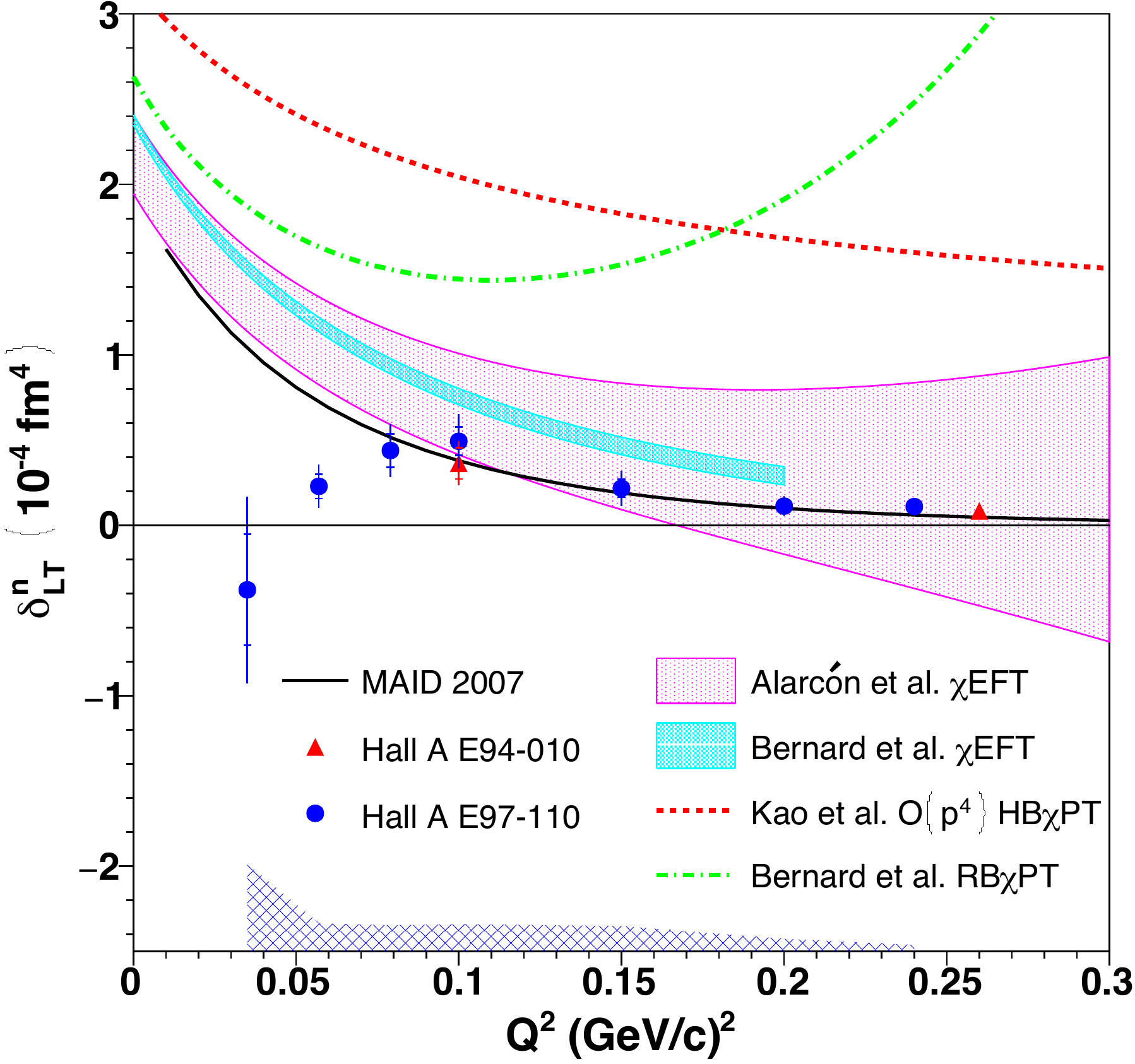}
    \includegraphics[width=0.40\textwidth]{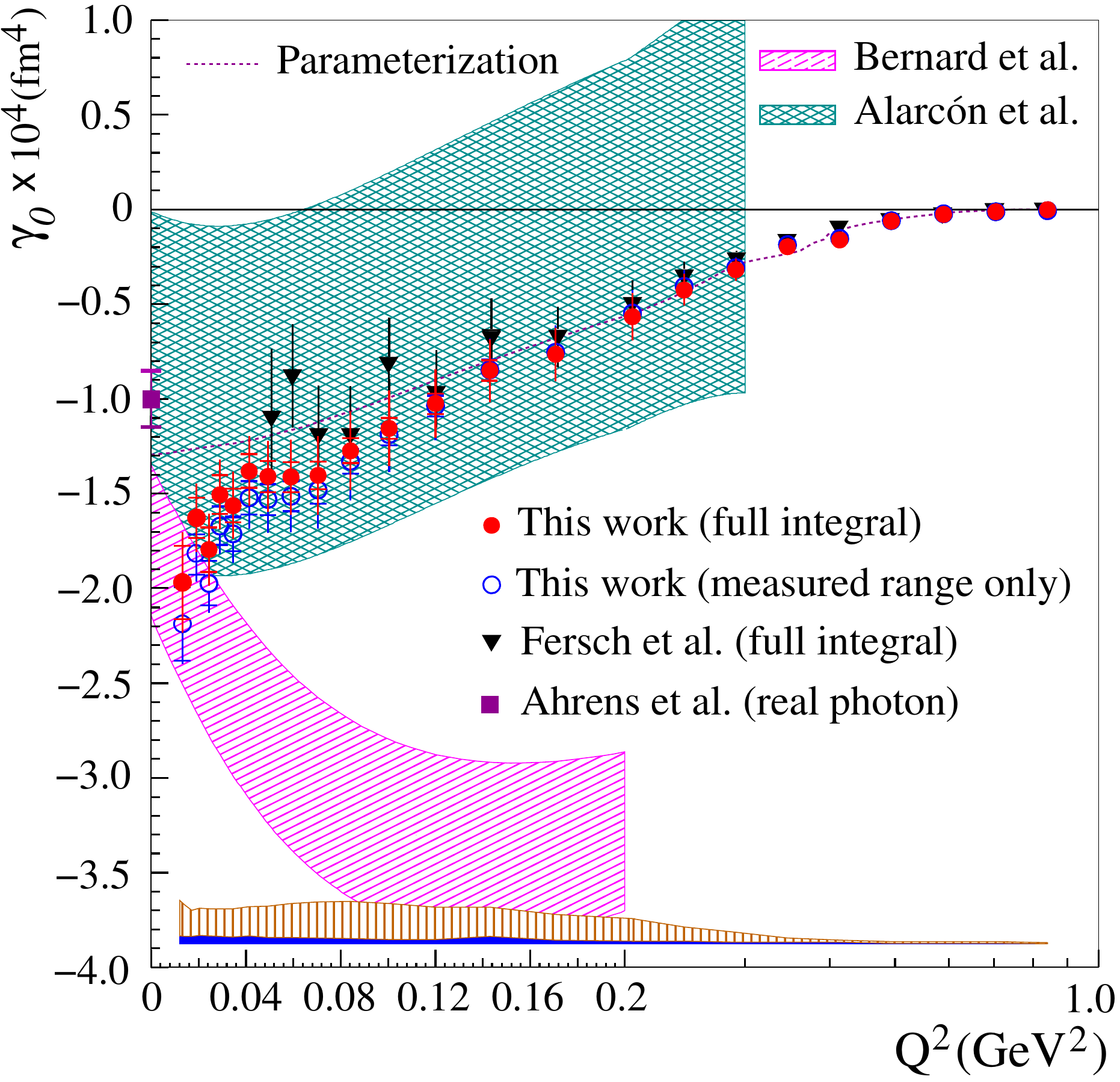}
    \includegraphics[width=0.40\textwidth]{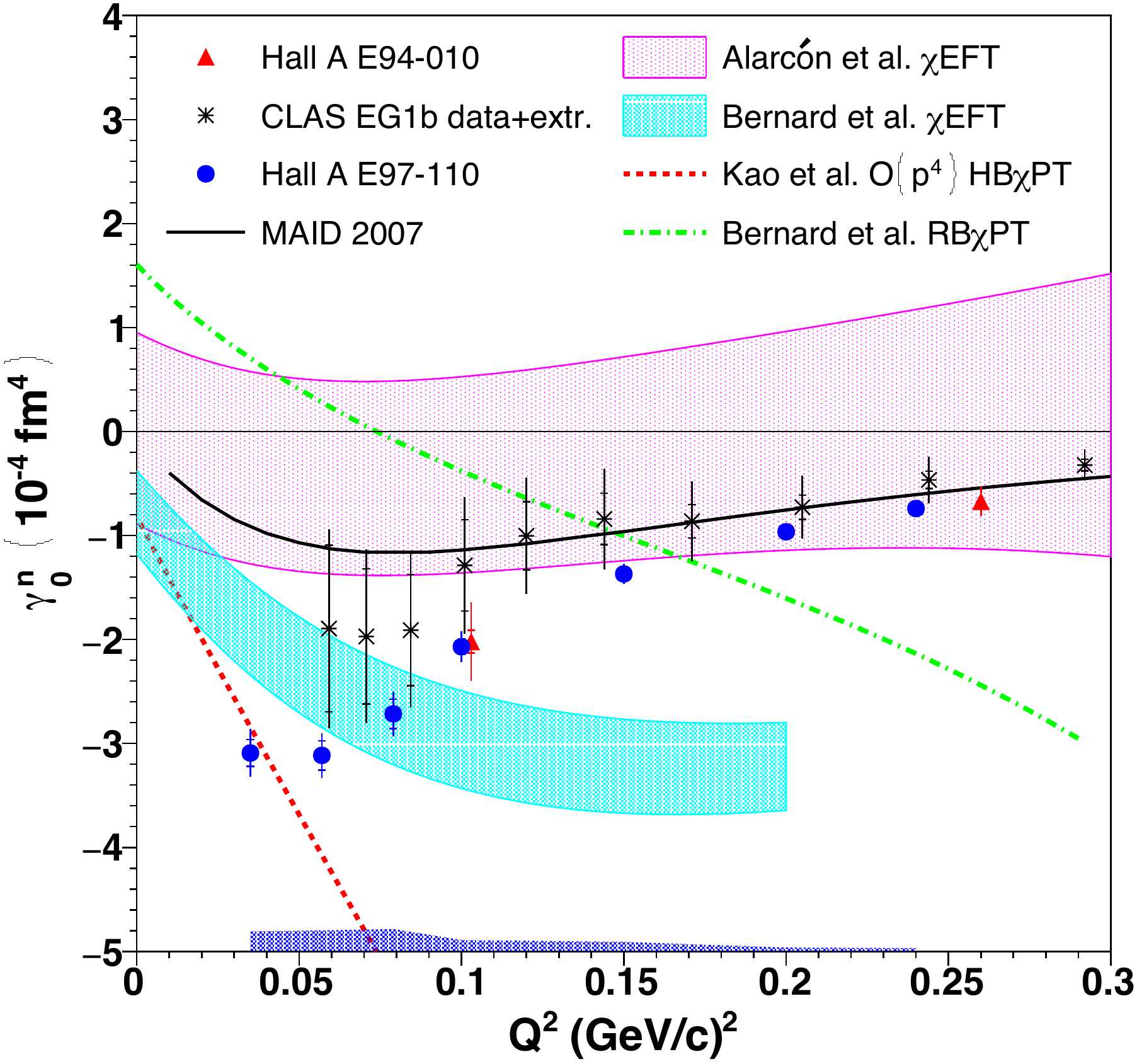}
    \includegraphics[width=0.40\textwidth]{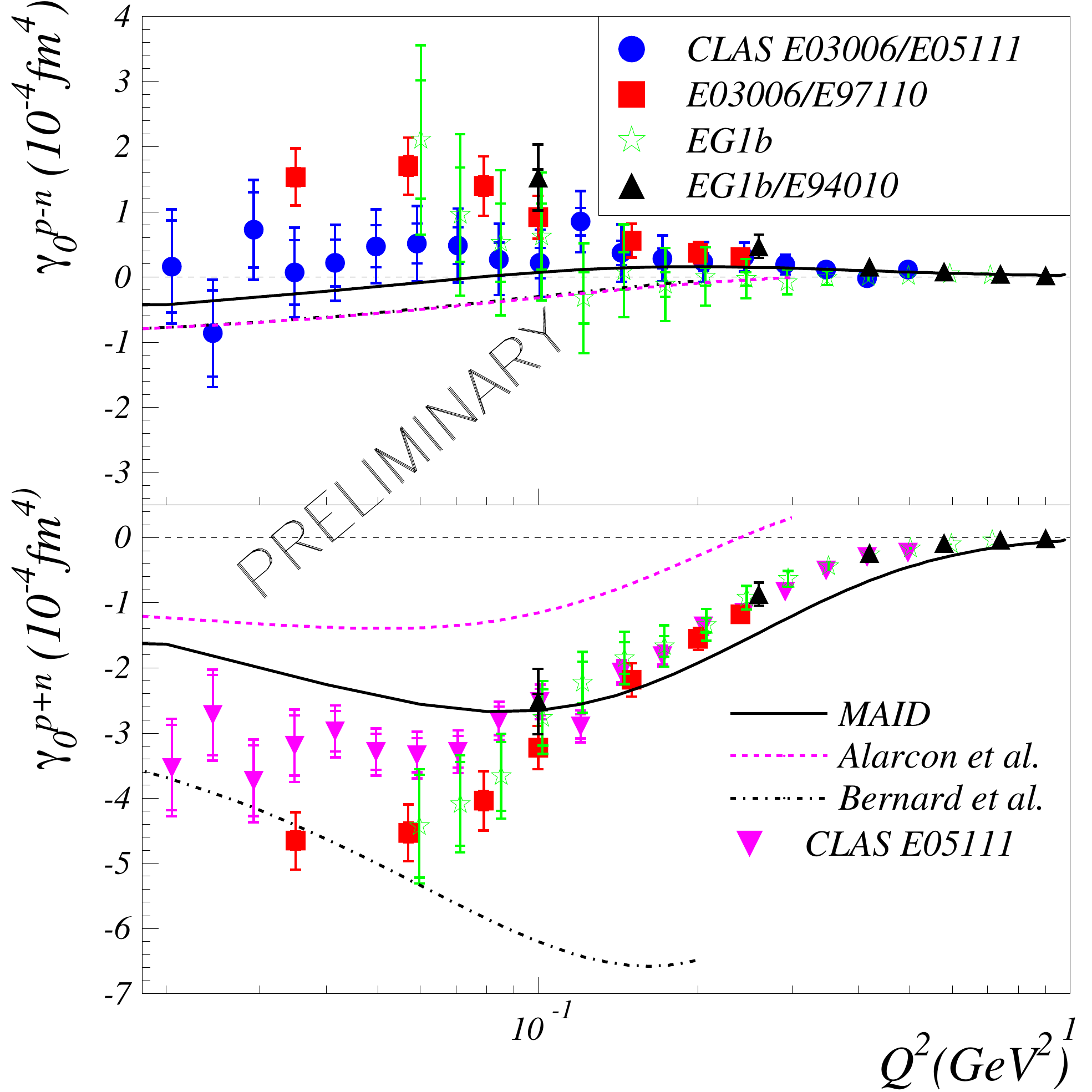} 
    \vspace{-3mm}
    \caption{\label{Fig:polarizabilities} 
Data on the generalized spin polarizabilities $\delta_\mathrm{LT}^n(Q^2)$ (top left), $\gamma_0^p(Q^2)$ 
 (top right), $\gamma_0^n(Q^2)$  (bottom left) and the isospin  decomposition of $\gamma_0(Q^2)$  (bottom right).
Inner error bars, sometimes too small to see, give the statistical uncertainties. 
Outer error bars are the quadratic sum of statistical  and uncorrelated systematic uncertainties. 
The horizontal bands give the correlated systematic uncertainties.
Also shown are state-of-the-art $\chi$EFT calculations~\cite{Bernard:2012hb, Lensky:2014dda} (Alarc\'on {\it et al.} $\chi$EFT, 
Bernard {\it et al.} $\chi$EFT), earlier calculations~\cite{Kao:2002cp, Bernard:2002bs} 
(Kao {\it et al.} HB$\chi$PT, Bernard {\it et al.} RB$\chi$PT)
and results from the  MAID model.}
\end{figure}

Results on the generalized spin polarizabilities $\gamma_0^p(Q^2)$~\cite{Zheng:2021yrn}, 
$\gamma_0^n(Q^2)$~\cite{Sulkosky:2021qmh}, their isospin decomposition $\gamma_0^{p\pm n}(Q^2)$,  
and $\delta_\mathrm{LT}^n(Q^2)$~\cite{Sulkosky:2021qmh} are shown in Fig.~\ref{Fig:polarizabilities}.

The $\delta_\mathrm{LT}^n$ data from E97110  agree well with the earlier E94010 data. At $Q^2 \gtrsim0.08$ GeV$^2$, 
the data agree with the latest $\chi$EFT calculations~\cite{Bernard:2012hb, Lensky:2014dda}
and the phenomenological MAID model~\cite{Drechsel:1998hk}. Data and predictions disagree at lower $Q^2$ despite the fact
that there, $\chi$EFT  calculations should be most robust.
Therefore, while the agreement between the latest $\chi$EFT calculations and E94010 data suggested that the $\delta_\mathrm{LT}$
puzzle may be solved, the new data refute it.
The renewed surprise with $\delta_\mathrm{LT}^n$ makes it interesting  to investigate the
integral $I_\mathrm{LT}(Q^2)$~\cite{Schwinger:1975ti} since it involves the same integrand as $\delta_\mathrm{LT}$ but without the $x^2$ weighting:
\begin{eqnarray}
I_\mathrm{LT}(Q^2)  =   \frac{8M^2}{Q^{2}}\int_0^{x_0}\Bigl[g_1(x,Q^2)+g_2(x,Q^2)\Bigr]dx.
\label{eq:I_LT SR}
\end{eqnarray}
The Schwinger sum rule  gives 
$I_\mathrm{LT}(Q^2)  \xrightarrow [Q^2 \to 0]~ \kappa e $, with $e$ the target particle electric charge i.e. $e=0$ for the neutron.
$I_\mathrm{LT}^n$ is shown on the bottom right panel of Fig.~\ref{Fig:1st_mom}. 
The sum rule expectation $I_\mathrm{LT}^n(0)=0$ agrees with the E97-110 data once they are guided to $Q^2=0$ using the
expected behavior of $I_\mathrm{LT}(Q^2)$ from the Gerasimov-Drell-Hearn (GDH) sum rule~\cite{Gerasimov:1965et} 
and elastic form factors. 
Since the GDH sum rule is solid, with its validity verified to good accuracy~\cite{Strakovsky:2022tvu}, 
and the form factors are well measured, 
the agreement strengthens confidence in the data quality. Yet, it cannot entirely rule out possible issues with the low-$x$
extrapolation 
(important in $I_\mathrm{LT}$ while suppressed in $\delta_\mathrm{LT}$) or
high-$x$ contamination from elastic/quasi-elastic reactions (enhanced in $\delta_\mathrm{LT}$
compared to $I_\mathrm{LT}$). The former is unlikely because the low-$x$ issue would need to conspire with a problem
in the data themselves so that $I_\mathrm{LT}$, $\Gamma_1$ and $\Gamma_2$ (Fig.~\ref{Fig:1st_mom}) still
agree with their expectations. A large-$x$ contamination would have to be mild enough so that  $I_\mathrm{LT}$,
 $\Gamma_1$ and $\Gamma_2$ still conform to expectations but important enough so that $\delta_\mathrm{LT}$ does not. 
\\
The E03006 data on $\gamma_0^p$ agree well with the earlier EG1 data (Fersch {\it et al.}) and
 the $\chi$EFT result of Alarc\'on {\it et al.} They agree with that of 
Bernard {\it et al.} only for the lowest $Q^2$ points. 
Also shown in the top right panel of Fig.~\ref{Fig:polarizabilities} is the datum at $Q^2=0$ from MAMI~\cite{Dutz:2004zz}. 
At first, it may seem incompatible with the E03006 data but their extrapolation to $Q^2=0$ assuming either
the results of Bernard {\it et al.} or Alarc\'on {\it et al.} shows that under this assumption, the data of JLab and MAMI agree within uncertainties~\cite{Strakovsky:2022tvu}.\\
The Hall A E97110 data on $\gamma_0^n$ agree with the earlier EG1 (Hall B) and E94010 (Hall A) data, 
but not with the predictions except at higher $Q^2$ for Alarc\'on {\it et al.} and MAID. 
Also shown are the older $\chi$EFT calculations~\cite{Bernard:2002bs, Kao:2002cp}.\\
The bottom right panel of  Fig.~\ref{Fig:polarizabilities} shows the isospin decomposition  of $\gamma_0$. The 
new and previous data agree. The E03006/E97110 and E03006/E05111 combinations  agree 
with each other, but there is a tension at lower $Q^2$. The two combinations differ in the origin of their neutron 
information (from $^3$He for E03006/E97110 and D for E03006/E05111) but also in the proton one
since the proton presents in D affects both $\gamma_0^{p\pm n}$
quantities: p$-$n$\simeq$2p$-$D and p$+$n$\simeq$D. 
The two combinations suggest that $\gamma_0^{p-n}$ remains
positive in the $Q^2$ domain experimentally covered, in contrast to the $\chi$EFT and MAID predictions. For $\gamma_0^{p+n}$,  
both combinations agree with Bernard {\it et al.} for the lowest $Q^2$ points, and disagree  with Alarc\'on {\it et al.} and MAID.
 
\section{Results on first moments}
\begin{figure}[!h]
  \centering
    \includegraphics[width=0.42\textwidth, height=0.34\textwidth]{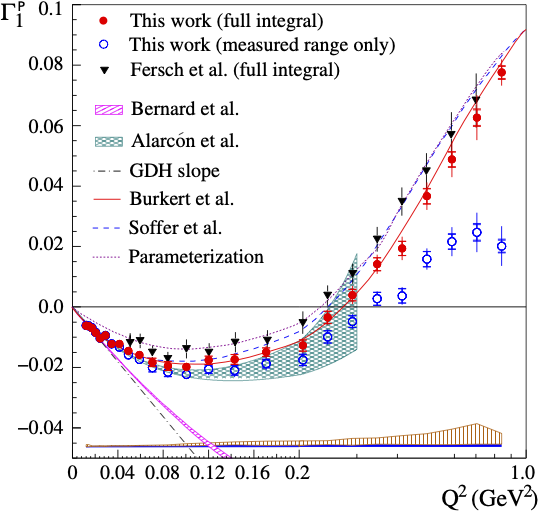}
    \includegraphics[width=0.46\textwidth, height=0.34\textwidth]{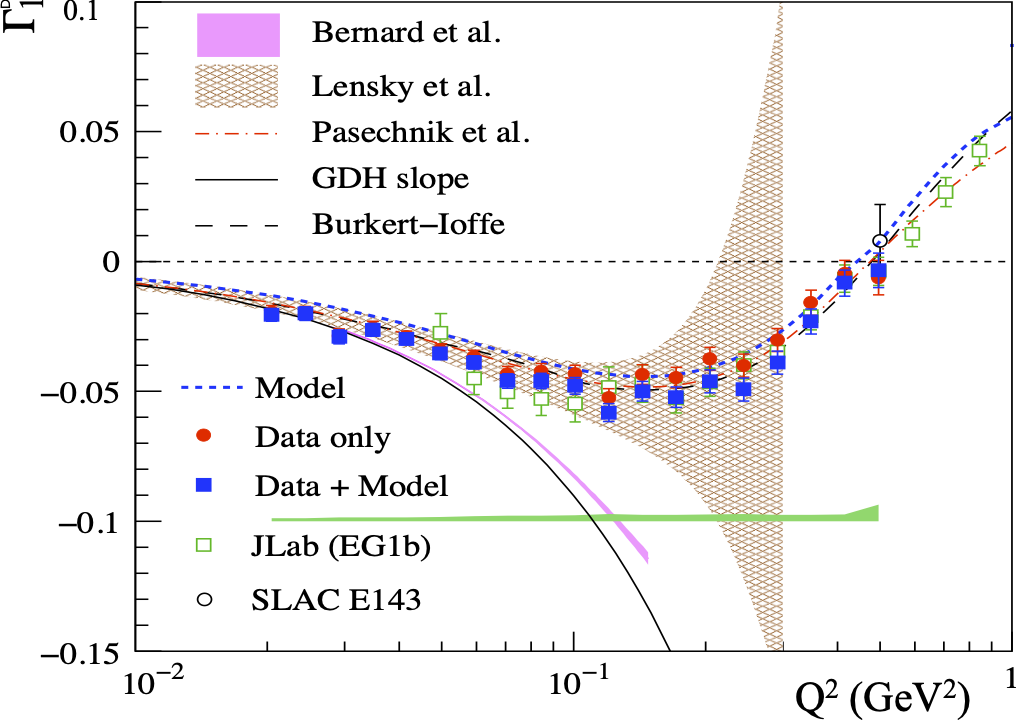} 
    \includegraphics[width=0.45\textwidth, height=0.37\textwidth]{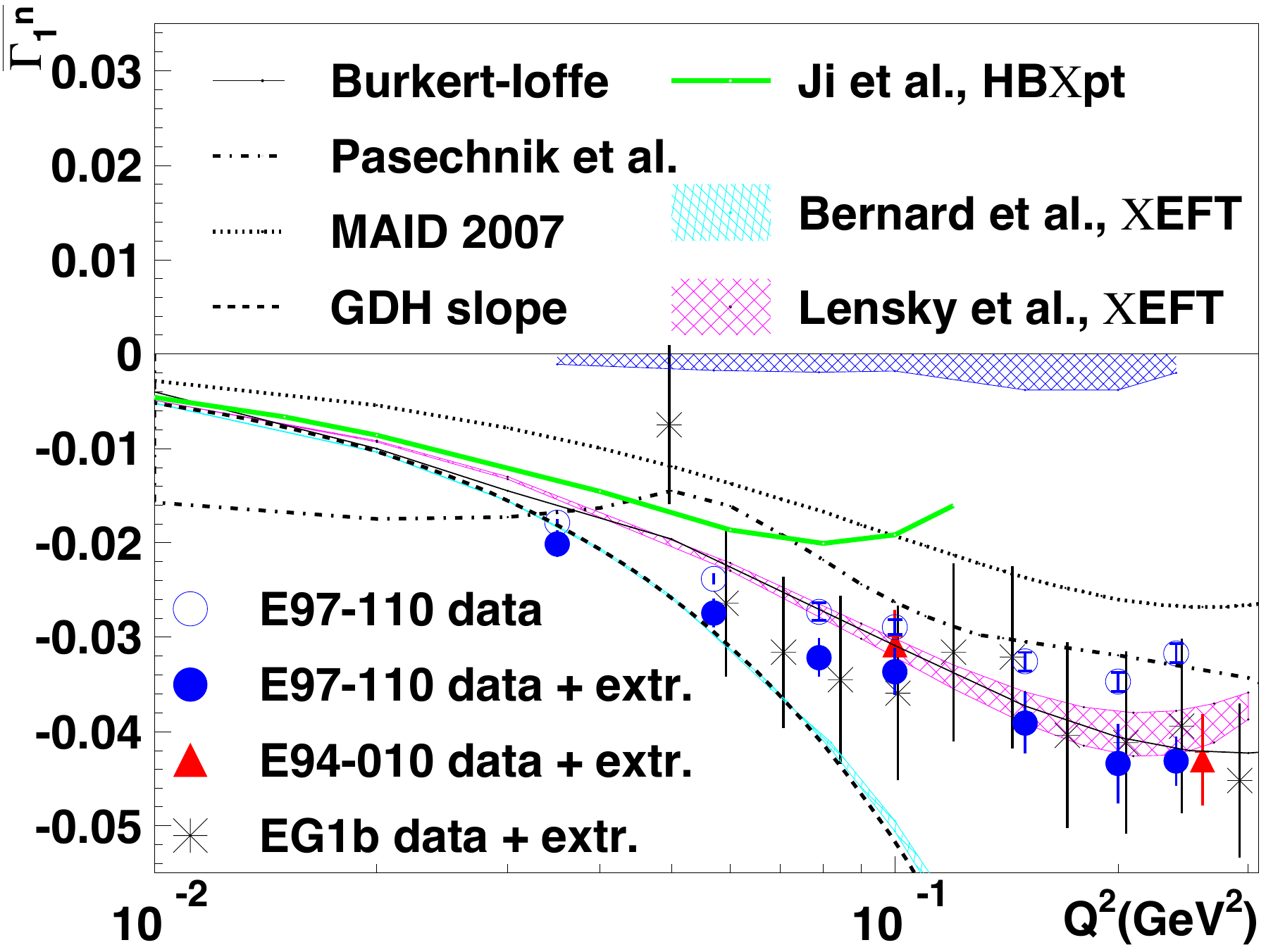}
    \includegraphics[width=0.42\textwidth, height=0.4\textwidth]{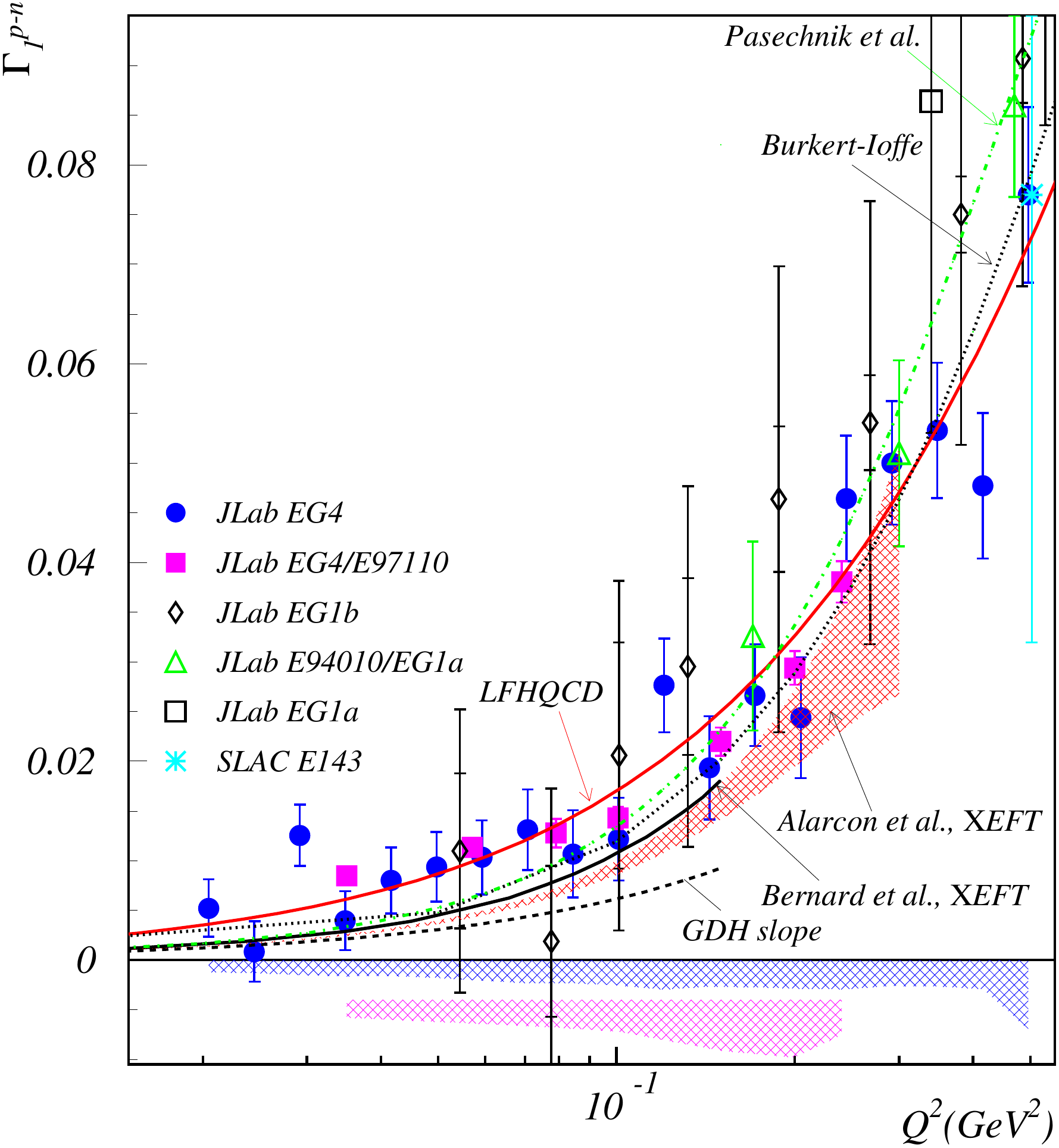}  
    \includegraphics[width=0.42\textwidth, height=0.37\textwidth]{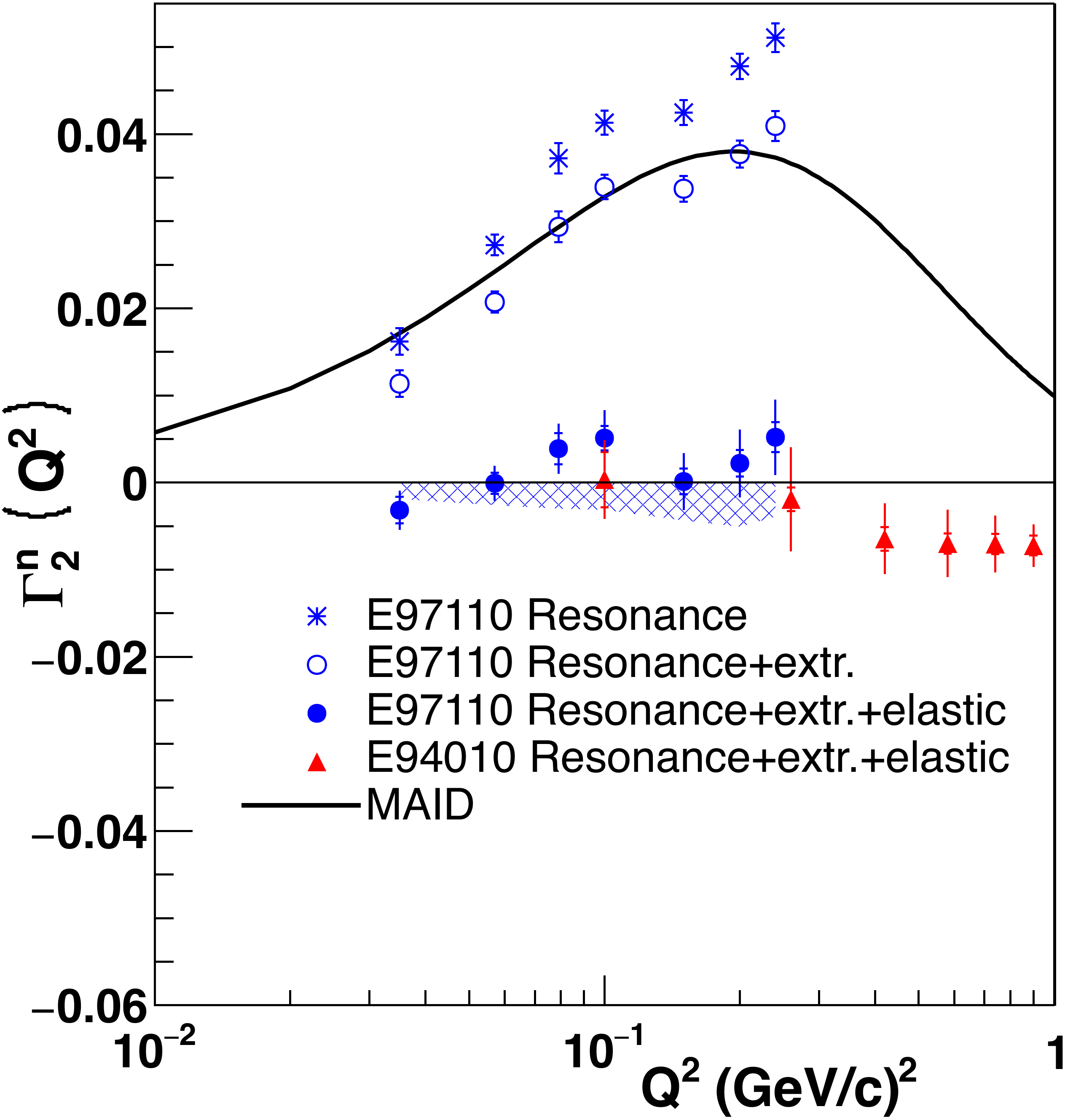} 
     \includegraphics[width=0.46\textwidth, height=0.37\textwidth]{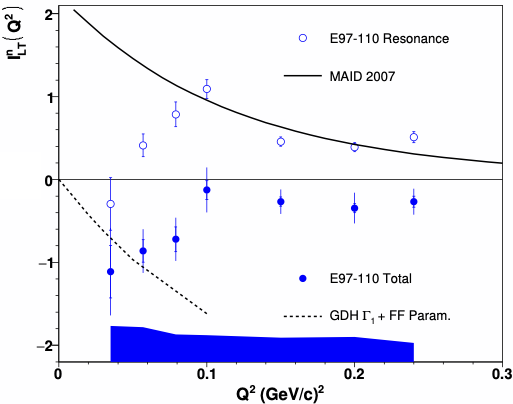} 
    \vspace{-3mm}
    \caption{\label{Fig:1st_mom} 
 The first moments $\Gamma_1$ 
 (first 4 panels), $\Gamma_2^n$ (bottom left) and $I_\mathrm{LT}^n$ (bottom right).}
\end{figure}

The first moment $\Gamma_1 \equiv \int g_1 dx$ is shown in Fig.~\ref{Fig:1st_mom} for the proton, neutron, deuteron and the Bjorken sum $\Gamma_1^{p-n}$. First moments $I_\mathrm{LT}^n$ and   
$\Gamma_2^n \equiv \int g_2^n dx$ are also shown.
The same observations as for  $\gamma_0^p$ stand for $\Gamma_1^p$: the 
new data agree well with the earlier EG1 data and with the latest $\chi$EFT calculations, 
albeit  only for the lowest  $Q^2$ points for Bernard {\it et al.} 
The same holds for $\Gamma_1^D$.
 The E97110 data on $\Gamma_1^n$ agree reasonably with the latest $\chi$EFT calculations, in contrast to $\gamma_0^n$.
 The new Hall A data agree well with the earlier Hall B (EG1) and Hall A  (E94010) data.\\
 The combinations E03006/E05111 and E03006/E97110 used to form the Bjorken sum $\Gamma_1^{p-n}$ 
 agree  with each other and with the earlier data~\cite{Deur:2004ti, Deur:2008ej, Abe:1998wq}. 
 They also agree qualitatively with  the predictions from $\chi$EFT  and the several
 available models~\cite{Burkert:1992tg, Brodsky:2010ur} but the predictions are systematical larger at low $Q^2$ 
 (except the LFHQCD calculation
~\cite{Brodsky:2010ur}). 
 This makes the $\Gamma_1^{p-n}=bQ^2+cQ^4$ fit to the data, performed to provide quantitative comparisons, 
 to yield a parameter $b$ larger than the GDH expectation and a $c$ (the $\chi$EFT prediction {\it per se}) of sign opposite 
 to the $\chi$EFT expectations~\cite{Deur:2021klh}. \\
The E97110 data for $\Gamma_2^n \equiv \int_0^1 g_2^n dx$ (where in contrast to other moments, the $x=1$ elastic contribution is
 included in the integral) agree well with the earlier data and with the Burkhardt-Cottingham sum rule expectation
 that $\Gamma_2=0$~\cite{Burkhardt:1970ti}.
 For $I_\mathrm{LT}^n$, see discussion in the previous section.

 \section{Conclusion}
 We can revisit Table~\ref{tab:comp_old}, adding to it the refined $\chi$EFT calculations and the new data at lower 
 $Q^2$ and of improved precision. 
\begin{table*}[!h]
\resizebox{0.999\textwidth}{!}{%
\begin{tabular}{|c|c|c|c|c|c|c|c|c|c|c|} \hline
~ & $\Gamma_1^p$~[3, 4]    &  $\Gamma_1^n$~[4, 5] & $\Gamma_1^{p-n}~[6, 7]$  & $\Gamma_1^{p+n}~[4, 7]$ & $\gamma_0^{p}~[4]$ &$\gamma_0^{n}~[8]$ &$\gamma_0^{p-n}~[7]$ &$\gamma_0^{p+n}~[4, 7]$ &  $\delta_\mathrm{LT}^n~[8]$ \\ \hline 
Ji {\it et al.}~\cite{Ji:1999pd} & {\bf \color{red}{X}} & {\bf \color{red}{X}} & {\bf \color{blue}{A}} & {\bf \color{red}{X}} & {\bf -} & {\bf -}  & {\bf -}  & {\bf -}  & {\bf -}  & {\bf -}\\ \hline
Bernard {\it et al.}~\cite{Bernard:2002bs} & {\bf \color{red}{X}} & {\bf \color{red}{X}} & {\bf \color{blue}{A}} & {\bf \color{red}{X}} & {\bf \color{red}{X}} & {\bf \color{blue}{A}}& {\bf \color{red}{X}}  & {\bf \color{red}{X}} & {\bf \color{red}{X$^*$}}  &{\bf \color{red}{X}}\\ \hline 
Kao {\it et al.}~\cite{Kao:2002cp} & {\bf -}  &{\bf -}  & {\bf -}  &{\bf -}   & {\bf \color{red}{X}} & {\bf \color{blue}{A}}& {\bf \color{red}{X}}  & {\bf \color{red}{X}} & {\bf \color{red}{X$^*$}} &{\bf \color{red}{X}}\\ \hline 
Bernard {\it et al.}~\cite{Bernard:2012hb} & {\bf \color{red}{X}} & {\bf \color{red}{X}} & {\bf \color{blue}{$\sim$A}} & {\bf \color{red}{X}} & {\bf \color{red}{X}} & {\bf \color{blue}{A}}& {\bf \color{red}{X$^*$}}  & {\bf \color{red}{X$^*$}} & {\bf \color{red}{X$^*$}}  &{\bf \color{red}{X}}\\ \hline 
Alarc\'on {\it et al.}~\cite{Lensky:2014dda} & {\bf \color{blue}{A}} & {\bf \color{blue}{A}} & {\bf \color{blue}{$\sim$A}} & {\bf \color{blue}{A}} & {\bf \color{blue}{$\sim$A}} & {\bf  \color{red}{X}}& {\bf \color{red}{X$^*$}}  & {\bf \color{red}{X$^*$}} & {\bf \color{blue}{A$^*$}}  &{\bf \color{red}{X}}\\ \hline 
\end{tabular}}
\vspace{-3mm}
\caption{Same as Table~\ref{tab:comp_old}  but including the latest data and $\chi$EFT results. The $^*$ denotes
preliminary data and {\bf \color{blue}{$\sim$A}} either an approximate agreement or an agreement over a range significantly smaller
than $Q^2<0.1$~GeV$^2$. 
}
\label{tab:comp_new} 
\end{table*}
The $\delta_\mathrm{LT}^p$ preliminary data refer to those of JLab Hall A E08027 (see K. Slifer and D. Ruth contributions
to these proceedings).
An advance is that all the observables in the table are now predicted by $\chi$EFT. Furthermore, 
there is a better agreement between data and predictions than in the past. Yet, puzzles remain. 
While for the Bjorken sum $\Gamma_1^{p-n}$ there is qualitative agreement, which was expected since the $\Delta_{1232}$ 
is suppressed in $\Gamma_1^{p-n}$~\cite{Burkert:2000qm}, there are disagreements for $\delta_\mathrm{LT}$ and $\gamma_0^{p-n}$
where the $\Delta_{1232}$ is also suppressed and have the additional  advantage that as
higher moments, they have little missing low-$x$  contribution. On the other hand, a complication with $\gamma_0$ and $\delta_\mathrm{LT}$ is that their value and slope at $Q^2=0$ must be calculated. In contrast,  $\Gamma_1(0)$ is known as it {\it must} vanish, and its slope is given by the GDH sum rule.

Therefore, despite its success in many domains, $\chi$EFT  remains challenged by nucleon spin observables, 
the latest data coming from 
dedicated low $Q^2$ experiments. Low $Q^2$ sum rule measurements are undeniably challenging:
forward angle detection is difficult to reach and subjected to large backgrounds, 
a large $\nu$ range is needed, 
there are low-$x$ extrapolations, 
avoiding high-$x$ contamination requires a careful analysis... 
An additional challenge for neutron data is that nuclear corrections are needed and, while the general 
agreement between the neutron data coming from deuteron and $^3$He is encouraging, one may ask 
how reliable the corrections are at low $Q^2$.
Yet, the experiments -- old and new -- provide consistent results and conclusion while being independent and having quite different detectors 
and methods. 
One must note the disagreement between the state-of-the-art $\chi$EFT predictions. But
it does not necessarily indicate an inconsistency. Rather, the difference in $Q^2$-behaviors seems to mostly arise from
including~\cite{Lensky:2014dda} or not~\cite{Bernard:2012hb}
phenomenological estimates of higher order terms of the $\chi$EFT series. In the case  where the value of the 
observable at $Q^2=0$ is not known (e.g. for $\gamma_0$ or $\delta_\mathrm{LT}$), an additional important difference 
comes from enforcing~\cite{Lensky:2014dda} or not~\cite{Bernard:2012hb} consistent couplings. 
Therefore, it remains unclear what the origin of the experiment/theory discrepancy is. 
A possibility to advance further, even if no new
experiments measuring these observables are foreseen and calculating the next order of the $\chi$EFT series
is very difficult, is to compute the observables with other non-perturbative approaches, e.g.
that based on the Dyson-Schwinger Equations, Lattice QCD, Gauge-Duality (AdS/QCD)  or global 
phenomenological analyses like MAID or SAID~\cite{Strakovsky:2022tvu}.
It is important to resolve this issue: it  challenges our search for a description of Nature at all level since 
$\chi$EFT is the leading approach to manage the first level of complexity arising above 
the strong force sector of the Standard Model. 

~

\noindent {\bf Acknowledgements} This material is based upon work supported by the U.S. Department of Energy, Office of Science, Office of Nuclear Physics under contract DE-AC05-06OR23177.

\end{document}